\documentclass[twocolumn,showpacs,preprintnumbers]{revtex4}
\usepackage{amsmath}
\usepackage{dcolumn}
\usepackage{bm}
\usepackage{graphicx}
\usepackage{amsfonts}
\usepackage{amssymb}
\usepackage{mathrsfs}
\usepackage{color}

\begin{document}

\title{Parity-relevant Zitterbewegung and quantum simulation by a single trapped ion}
\author{Kunling Wang$^{1, 2}$, Tao Liu$^{3}$, Mang Feng$^{1}$\footnote{E-mail: mangfeng@wipm.ac.cn},
Wanli Yang$^{1}$ and Kelin Wang$^{4}$}
\affiliation{$^{1}$ State Key Laboratory of Magnetic Resonance and
Atomic and Molecular Physics, Wuhan Institute of Physics and
Mathematics, Chinese Academy of Sciences, Wuhan, 430071, China\\
$^{2}$ Graduate School of the Chinese Academy of Sciences, Beijing
100049, China \\ $^{3}$ The School of Science, Southwest University
of Science and Technology, Mianyang 621010, China \\
$^{4}$ The Department of Modern Physics, University of Science and
Technology of China, Hefei 230026, China}


\begin{abstract}
Zitterbewegung (ZB), the trembling of free relativistic electrons in
a vacuum could be simulated by a single trapped ion. We focus on the
variations of ZB under different parity conditions and find no ZB in
the case of odd or even parity. ZB occurs only for admixture of the
odd and even parity states. We also show the similar role played by
the parity operator for the trapped ion in Fock-state representation
and the space inversion operator for a realistic relativistic
electron. Although the ZB effect is invisible in a relativistic
electron, preparation of the trapped ion in different parity states
is a sophisticated job, which makes it possible to observe the parity
relevant ZB effects with currently available techniques.
\end{abstract}

\maketitle


Since the discovery by Schr$\ddot{o}$dinger \cite {sch},
Zitterbewegung (ZB), i.e., the trembling of free relativistic
electrons, has drawn more and more attention and interests over past
years \cite{p0,p1,p2,p3,p5,p4,p6}. It has been generally believed
that the trembling of a relativistic electron is resulted by the
interference between negative and positive energy components, a
typical relativistic feature of the Dirac electron. Up to now,
however, no direct observation of ZB has been achieved due to
inaccessibility with current experimental techniques, which led to
some questioning on the ZB \cite {p2,ques}. Moreover, there have
been alternative explanations for the origin of the ZB, such as the
continuously virtual transition process between different internal
states in view of quantum field theory \cite{p3} or the relevance to
the complex phase factor in context of space-time algebra
\cite{p6}.

By quantum simulation, some relativistic effects have recently been
demonstrated in some controllable physical systems, such as
graphene, semiconductor, superconductor and trapped ion
\cite{s5,c,a,s1,s2,s3,s4,s7,s8}. It was recently considered that the
ZB occurs not only under the relativistic condition, but also
extensively in the dynamics of a system with more than one degree of
freedom \cite{s5,s9}. The demonstration of the ZB beyond the
relativistic electron helps us to further understand the Dirac
equation and the relevant relativistic phenomena.

In this work, we focus on the role of parity played in the
simulation of Dirac equation for the ZB effect by a single trapped
ion. Like in \cite {c,a}, we also employ the motional degrees of
freedom of the ion to simulate the position and momentum of the
relativistic electron, and the internal degrees of freedom to refer
to the energy states. Since the motional state of the ion could be
quantized, we may discuss the problem in number-state
representation. The key point of our work is to introduce a parity
operator $\hat{\Pi}$, by which we show that, besides the
conventional consideration of the origin of the ZB, i.e., the
interference between the positive- and negative-energy components,
the ZB is also relevant to parity of the states. To understand the
relevant physics, we will compare the parity operator of the trapped
ion with the space inversion operator of the realistic relativistic
electron. The experimental feasibility to observe the
parity-relevant ZB effects will be justified.

Under the radiation of three laser lights with red-detuning,
blue-detuning, and carrier transition, respectively, the interacting
Hamiltonian of a single trapped ion reads \cite{a,c}
\begin{equation} \label{eq:1}
H=i\hbar \eta \widetilde{\Omega} (a^+ - a) \hat{\sigma}_x + \hbar
\Omega \sigma_z,
\end{equation}
where $\eta$ is the Lamb-Dicke parameter, $\widetilde{\Omega}$ and
$\Omega$ are, respectively, the effective Rabi frequency and the
effective Larmor frequency of the ion. $a^{+}$ ($a$) is the creation
(annihilation) operator of the quantized motion of the ion.
$\sigma_x$ and $\sigma_z$ are usual Pauli operators. Defining $p=i
\hbar (a^+ - a)/2\Delta$ with $\Delta$ the size of the ground state
wave function, we may rewrite the Hamiltonian as a form analogous to
a $1 + 1$ dimensional Dirac equation
\begin{equation} \label{eq:2}
H_D=2\eta\Delta\widetilde{\Omega} p \sigma_x + \hbar\Omega\sigma_z
\end{equation}
provided $c:=2\eta\Delta\widetilde{\Omega}$ and $mc^2:=\hbar \Omega$.

Introducing the parity operator
\begin{equation} \label{eq:3}
\hat{\Pi}=e^{i \pi(a^+ a - \frac{1}{2}+\frac{1}{2}\sigma_z)}
\end{equation}
commuting with the Hamiltonian in Eq. (\ref{eq:1}), which means
$\hat{\Pi}$ is a conserved quantity under the Hamiltonian $\hat{H}$, we
study below the dynamics of the trapped ion with different parity
states, i.e., of the odd or even parity, or of admixture of the
both.

The ZB is related to the average position $\left\langle x(t)
\right\rangle$ of the trapped ion. Since our interest is in the
states under some parity conditions, we have to find the common
eigenfunctions of $\hat{H}$ and $\hat{\Pi}$. To this end, we define
$\left|p\right\rangle$ as the eigenfunction of $(a^{+} - a)$ by
assuming $(a^+ -a)\left|p\right\rangle= -ip\left|p\right\rangle$.
Straightforward deduction yields
\begin{equation} \label{eq:4}
\left|p\right\rangle=\frac{1}{\sqrt[4]{2\pi}}e^{-\frac{p^2}{4}}\sum^{\infty}_{n=0}\frac{i^n
H_n(p)}{\sqrt{n!}}\left|n\right\rangle,
\end{equation}
where $H_n(x)$ is a Hermite polynomial defined by $H_n(x)=(-1)^n
e^{x^2/2}\frac{d^n}{dx^n}e^{-x^2/2}$ \cite{2}. It could be proven
that the states in \{$\left|p\right\rangle$\} are orthonormal and
complete, and each state $\left|p\right\rangle$ corresponds to the
momentum $p_{mom}=\hbar p/(2\Delta)$.

Substituting Eq.(\ref{eq:4}) into Eq.(\ref{eq:1}) and diagonalizing $\hat{H}$ yields the eigenvalues
\begin{equation*}
E_{\pm}=\pm\sqrt{(\hbar\Omega)^2 + (\eta\hbar\widetilde{\Omega}p)^2}
\end{equation*}
and the eigenstates
\begin{eqnarray}
\left|S_{E_{+}}(p)\right\rangle=N(p)\left[1,
\frac{\eta\hbar\widetilde{\Omega}p}{\left|E_{+}\right|+\hbar\Omega}
\right]^{T}, \nonumber  \\
\left|S_{E_{-}}(p)\right\rangle=N(p)
\left[\frac{-\eta\hbar\widetilde{\Omega}p}{\left|E_{-}\right|+\hbar\Omega},
1 \right]^{T},   \nonumber
\end{eqnarray}
with $N(p)=\sqrt{(\left|E_{\pm}\right|+\hbar\Omega)/2\left|E_{\pm}\right|}$ the
normalization factor. So the total eigenfunctions of $\hat{H}$
are
$\left|\psi_{E_{+}}(p)\right\rangle=\left|S_{E_{+}}(p)\right\rangle\otimes\left|p\right\rangle$
and
$\left|\psi_{E_{-}}(p)\right\rangle=\left|S_{E_{-}}(p)\right\rangle\otimes\left|p\right\rangle$.

Under the parity operator $\hat{\Pi}$, we assume following parity
states in the number-state representation,
\begin{eqnarray} \label{eq:o}
\left|o\right\rangle=\left|+\right\rangle\sum_m f_{2m+1}\left|2m+1\right\rangle+\left|-\right\rangle
\sum_m f_{2m}\left|2m\right\rangle, \\
\left|e\right\rangle=\left|+\right\rangle\sum_m
f_{2m}\left|2m\right\rangle+\left|-\right\rangle\sum_m
f_{2m+1}\left|2m+1\right\rangle, \label{eq:e}
\end{eqnarray}
where $f_{k}$ ($k=2m$, $2m+1$, $m=0, 1, 2, \cdots$) are
normalized coefficients. $\left|o\right\rangle$ stands for odd
parity state due to
$\hat{\Pi}\left|o\right\rangle=-\left|o\right\rangle$ and
$\left|e\right\rangle$ for even parity state with
$\hat{\Pi}\left|e\right\rangle=\left|e\right\rangle$.
$\left|\pm\right\rangle$ are the eigenstates of $\sigma_z$ with
eigenvalues $\pm 1$, respectively.

For our purpose, we may construct the spin states
$\left|\pm\right\rangle$ by positive energy eigenstates as
\begin{eqnarray} \label{eq:e++}
\left|+\right\rangle = W (\left|S_{E+}(p)\right\rangle+\left|S_{E+}(-p)\right\rangle), \\
\left|-\right\rangle = W'
(\left|S_{E+}(p)\right\rangle-\left|S_{E+}(-p)\right\rangle),
\label{eq:e+-}
\end{eqnarray}
or by negative energy eigenstates as,
\begin{eqnarray} \label{eq:e-+}
\left|+\right\rangle = W' (\left|S_{E-}(-p)\right\rangle-\left|S_{E-}(p)\right\rangle), \\
\left|-\right\rangle = W
(\left|S_{E-}(p)\right\rangle+\left|S_{E-}(-p)\right\rangle),
\label{eq:e--}
\end{eqnarray}
with normalization factors
$W=\sqrt{\left|E_{\pm}\right|/(2\left|E_{\pm}\right|+2\hbar \Omega)}$
and $W'=\sqrt{[\left|E_{\pm}\right|(\left|E_{\pm}\right|+\hbar
\Omega)]/2(\eta\hbar\widetilde{\Omega}P)^2}$. Moreover, the odd
and even motional states are associated with the momentum
eigenstates $\left|p\right\rangle$, i.e.,
\begin{eqnarray} \label{eq:odd}
\sum_m f_{2m+1}\left|2m+1\right\rangle \propto \frac{1}{\sqrt{2}} (\left|p\right\rangle - \left|-p\right\rangle), \\
\sum_m f_{2m}\left|2m\right\rangle \propto \frac{1}{\sqrt{2}}
(\left|p\right\rangle + \left|-p\right\rangle). \label{eq:even}
\end{eqnarray}
Substituting Eqs. (\ref{eq:e++}), (\ref{eq:e+-}), (\ref{eq:odd}) and
(\ref{eq:even}) into Eq. (\ref{eq:o}), we may write down the
co-eigenstate of E$_{+}$ and odd parity to be,
\begin{eqnarray*}
\left|\psi^o_{E_{+}}\right\rangle =
W(p)(\left|S_{E_{+}}(p)\right\rangle+\left|S_{E_{+}}(-p)\right\rangle)
\otimes \frac{1}{\sqrt{2}}
(\left|p\right\rangle - \left|-p\right\rangle) \\
+ Q W'(p) (\left|S_{E_{+}}(p)\right\rangle - \left|S_{E_{+}}(-p)\right\rangle)\otimes (\frac{1}{\sqrt{2}} (\left|p\right\rangle
+ \left|-p\right\rangle) \\
\end{eqnarray*}
where Q is a coefficient to be determined. We only keep some reasonable
terms by setting $Q= W(p)/W'(p)$, i.e., elimination of the terms of
$\left|S_{E_{+}}(p)\right\rangle \left|-p\right\rangle$ and
$\left|S_{E_{+}}(-p)\right\rangle \left| p\right\rangle$. Then we
have,
\begin{equation} \label{eq:+o}
\left|\psi^o_{E_{+}}\right\rangle=\frac{1}{\sqrt{2}}\left[\left|\psi_{E_{+}}(p)\right\rangle-\left|\psi_{E_{+}}(-p)\right\rangle\right].
\end{equation}
Similarly, we have other co-eigenstates of E$_{+}$ and even parity,
E$_{-}$ and different parities \cite {explain},
\begin{eqnarray}
\left|\psi^e_{E+}\right\rangle=\frac{1}{\sqrt{2}}\left[\left|\psi_{E+}(p)\right\rangle+\left|\psi_{E+}(-p)\right\rangle\right],\\
\left|\psi^o_{E-}\right\rangle=\frac{1}{\sqrt{2}}\left[\left|\psi_{E-}(p)\right\rangle+\left|\psi_{E-}(-p)\right\rangle\right], \\
\left|\psi^e_{E-}\right\rangle=\frac{1}{\sqrt{2}}\left[\left|\psi_{E-}(p)\right\rangle-\left|\psi_{E-}(-p)\right\rangle\right].
\label{eq:-e}
\end{eqnarray}

Based on these eigenstates, the average position of the
ion $\left\langle x(t) \right\rangle=\left\langle (\hat{a}^+
+a)\Delta \right\rangle$ can be calculated by the evolved states
with odd or even parity, where
\begin{eqnarray*}
\left|\psi^o(t)\right\rangle=\sum_p{a_p\left|\psi^o_{E_{+}}(p)\right\rangle
e^{-\frac{i\left|E_{+}\right|t}{\hbar}}}+
\sum_p{b_p\left|\psi^o_{E_{-}}(p)\right\rangle e^{\frac{i\left|E_{-}\right|t}{\hbar}}}\\
\left|\psi^e(t)\right\rangle=\sum_p{a_p\left|\psi^e_{E_{+}}(p)\right\rangle
e^{-\frac{i\left|E_{+}\right|t}{\hbar}}}+
\sum_p{b_p\left|\psi^e_{E_{-}}(p)\right\rangle
e^{\frac{i\left|E_{-}\right|t}{\hbar}}},
\end{eqnarray*}
with $a_{p}$ and $b_{p}$ the coefficients determined by the initial
condition.

To calculate $\left\langle x(t) \right\rangle$, we first consider
$\left\langle dx/dt \right\rangle$, the average velocity,
$$\left\langle\frac{dx}{dt}\right\rangle=\frac{\Delta}{i\hbar}\left\langle \left[
\hat{a}^+ +\hat{a},\hat{H}_D
\right]\right\rangle=2\eta\Delta\widetilde{\Omega}\left\langle
\hat{\sigma}_x\right\rangle.$$ Taking the odd parity as an example,
we have
\begin{eqnarray} \label{eq:eig}
\left\langle \psi^o(t)\right|\hat{\sigma_x}\left|
\psi^o(t)\right\rangle =\sum_{p} a_{p}^* a_{p} \left\langle
\psi^o_{E_{+}}(t)\right|\hat{\sigma_x}\left|
\psi^o_{E_{+}}(t)\right\rangle \nonumber \\ + \sum_{p} b_{p}^* b_{p}
\left\langle \psi^o_{E_{-}}(t)\right|\hat{\sigma_x}\left|
\psi^o_{E_{-}}(t)\right\rangle \nonumber \\ + \sum_{p} a_{p}^* b_{p}
e^{\frac{2i\left|E_{\pm}\right|t}{\hbar}} \left\langle
\psi^o_{E_{+}}(t)\right|\hat{\sigma_x}\left|
\psi^o_{E_{-}}(t)\right\rangle \nonumber \\ + \sum_{p} b_{p}^* a_{p}
e^{\frac{-2i\left|E_{\pm}\right|t}{\hbar}} \left\langle
\psi^o_{E_{-}}(t)\right|\hat{\sigma_x}\left|
\psi^o_{E_{+}}(t)\right\rangle.
\end{eqnarray}
It is easy to check in Eq. (17) that
$\left\langle\psi^o(t)\right|\hat{\sigma_x}\left|
\psi^o(t)\right\rangle=0$ since every term is zero. As a result, $\left\langle
x(t)\right\rangle$ remains unchanged. Similarly we have
$\left\langle \psi^e(t)\right|\hat{\sigma_x}\left|
\psi^e(t)\right\rangle =0$. These results imply that a trapped ion
in an eigenstate of $\hat{\Pi}$ would be on average static.

Because Eq. (\ref{eq:eig}) involves both the positive and negative
energy components, the ZB should occur, according to the
conventional viewpoint, due to their interference. Our result,
however, presents that the ZB depends not only on the interference
between the positive and negative energy components, but also on
parity of the states.

\begin{figure}
\centering
\includegraphics[width=.47\textwidth]{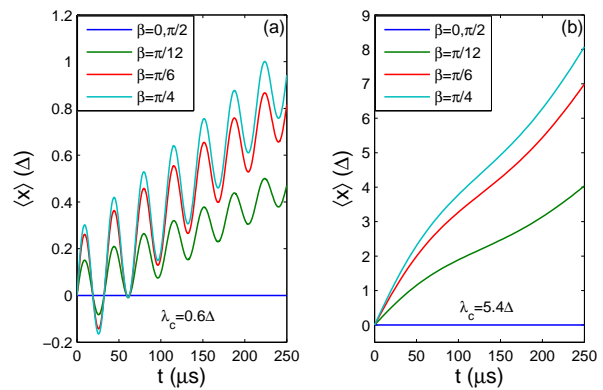}
\caption{(Color online) Numerical simulation of $\left\langle
\hat{x}(t)\right\rangle$ in units of $\Delta$ for initial state
$(\cos\beta|+\rangle + \sin\beta|-\rangle)|0\rangle$, where
$|\pm\rangle$ are eigenstates of $\sigma_{z}$ and $|0\rangle$ is the
ground motional state.
$\lambda_{c}=2\eta\tilde{\Omega}\Delta/\Omega$ is the Compton
wavelength. We consider (a) $\lambda_{c}=0.6\Delta$, near the
non-relativistic limit and (b) $\lambda_{c}=5.4\Delta$ close to the
relativistic limit. From the bottom to top, the curves correspond to
$\beta=0$ or $\pi/2$, $\pi/12$, $\pi/6$ and $\pi/4$, respectively.}
\label{fig:1}
\end{figure}

To be more clarified, we have numerically calculated in Fig.
\ref{fig:1} the average position of the trapped ion with the initial
state
$(\cos\beta\left|+\right\rangle+\sin\beta\left|-\right\rangle)\left|0\right\rangle$.
Since both $\left|+\right\rangle\left|n\right\rangle$ and
$\left|-\right\rangle\left|n\right\rangle$ are states with definite
parity, the case with $\alpha=\beta$ is of the most mixed parity. By
changing the values of $\beta$, we may see clearly from the figure
that the ZB occurs only in the admixture of the odd and even parity
eigenstates.

Our result could be understood by the viewpoint in \cite {s9} that
the ZB could appear in any system, besides the relativistic system,
with more than one degree of freedom. In our case, the ZB effect is
originated from the interplay between the internal and motional
degrees of freedom of the ion. With respect to the conventional
viewpoint of interference between the positive and negative energy
components, we may check Eq. (17) again which involves both positive
and negative energy states. The internal-motional-state interplay
leads to the interference between different energy component terms.
Once the ion is in a certain parity state, however, the interference
is destructive, yielding $\langle x\rangle =0$. So the ZB appears
only in the case of admixture of different parity states, which
allows the constructive interference between different energy
components. Simply speaking, whether the ZB appears or not, is
decided by both the interference and symmetry, the latter of which
is reflected by parity.

For a realistic relativistic electron, there is no demand to
quantize the motional freedom, but the parity operator $\hat{\Pi}$
discussed above reminds us of the space inversion operator $\hat{P}$
defined as \cite{l},
\begin{equation} \label{eq:P}
\hat{P}\varphi(\bar{x},t) = \hat{\sigma}_z\varphi(-\bar{x},t),
\end{equation}
where $\varphi(\bar{x},t)$ is the wavefunction of the relativistic
electron, and we have used the overline to represent the parameters
of the relative electron in order to distinguish from the ones of
the trapped ion. For a relativistic electron, the eigenstates of
$\hat{H}_{D}$ are,
\begin{eqnarray*}
\varphi_{\bar{E}_{+}}(\bar{p};\bar{x},t) = N(\bar{p})
\begin{pmatrix}  1 \cr
\frac {c\bar{p}}{|\bar{E}_{+}| + \bar{m}c^2}
\end{pmatrix}
e^{\frac{i\bar{p}\bar{x}}{\hbar}-\frac{i\left|\bar{E}_{+}\right|t}{\hbar}} \\
\varphi_{\bar{E}_{-}}(\bar{p}; \bar{x}, t) = N(\bar{p})
\begin{pmatrix}  \frac {-c\bar{p}}{|\bar{E}_{-}| + \bar{m}c^2} \cr
1
\end{pmatrix}
e^{\frac{i\bar{p}\bar{x}}{\hbar} + \frac{i\left|\bar{E}_{-}\right|t}{\hbar}},
\end{eqnarray*}
where $\bar{m}$ is the mass of the electron, $\bar{p}$ and $\bar{x}$
are, respectively, the momentum and the position, $\bar{E}_{\pm}$ stand
for the energies and N$(\bar{p})$ the normalization factor.

We have $\hat{P}\varphi_{\bar{E}_{\pm}}(\bar{p}; \bar{x},t) =
\pm\varphi_{\bar{E}_{\pm}}(-\bar{p}; \bar{x},t)$. Under such a
consideration, we may easily find the odd parity states
\begin{eqnarray*}
\varphi^o_{\bar{E}_{+}}=\frac{1}{\sqrt{2}}[\varphi_{\bar{E}_{+}}(\bar{p})-\varphi_{\bar{E}_{+}}(-\bar{p})], \\
\varphi^o_{\bar{E}_{-}}=\frac{1}{\sqrt{2}}[\varphi_{\bar{E}_{-}}(\bar{p})+\varphi_{\bar{E}_{-}}(-\bar{p})],
\end{eqnarray*}
and the even parity states
\begin{eqnarray*}
\varphi^e_{\bar{E}_{+}}=\frac{1}{\sqrt{2}}[\varphi_{\bar{E}_{+}}(\bar{p})+\varphi_{\bar{E}_{+}}(-\bar{p})], \\
\varphi^e_{\bar{E}_{-}}=\frac{1}{\sqrt{2}}[\varphi_{\bar{E}_{-}}(\bar{p})-\varphi_{\bar{E}_{-}}(-\bar{p})].
\end{eqnarray*}
They are formally consistent with the co-eigenstates Eqs.
(\ref{eq:+o})-(\ref{eq:-e}) for the trapped ion. In this sense, we
consider that $\hat{\Pi}$ and $\hat{P}$ play the same role in the
respective systems. In other words, $\hat{\Pi}$ is something like an
inversion parity operator in the number state representation, which
moves population back and forth between the internal and motional
states.

Following the same steps as for the trapped ion, we can immediately
find $\left\langle \bar{x}\right\rangle=0$ of the relativistic
electron under definite parity states. So the ZB of the relativistic
electron occurs when the electron is in both the admixture of
different parity states and the admixture of different energy
components. A particle, such as a relativistic electron or a trapped
ion, staying in a certain parity state will be static or in harmonic
oscillation. Otherwise, a moving particle with different energy
components coexisting will surely experience the ZB.

Since the ZB in a real relativistic electron is inaccessible with
current technique, we have to resort to other systems, such as the
trapped ion, to observe the parity-relevant ZB effect. With
different trapped ion's motional states, the eigenstates of the
parity operator have abundant forms, such as $\left|\pm\right\rangle
\otimes \left|n\right\rangle$ and $\left|+\right\rangle \otimes
\left|n\right\rangle_A \pm \left|-\right\rangle \otimes
\left|n\right\rangle_B$ where $\left|n\right\rangle_{A(B)}$ is the
displaced coherent state \cite{TLiu}.  With currently available
ion-trap technique, it has already been achieved the cooling of the
single ion to the motional ground state \cite {cooling}. Preparation
of the ion to a certain motional Fock state or a certain coherent
state has also been a sophisticated job \cite{fock}. As a result,
using the parity states mentioned above, we may check the variation
of the ZB with respect to different parity conditions following the
operations in \cite {c}.

Besides the simulation of Dirac equation, the parity operator
$\hat{\Pi}$ also has application in quantum computing. $\hat{\Pi}$
commutes not only with $\hat{H}$, but also with other laser-ion
interaction Hamiltonians, such as $\hat{H_r}=\sigma_+a+\sigma_-a^+$,
a usually used Hamiltonian for logic gate operation \cite {cz}. If
we prepare the ion in the state
$\left|+\right\rangle\left|0\right\rangle$ or other parity states,
the ion will be more steady during the operation than in any mixed
parity states. This is advantageous to quantum gate operation.

In summary, we have investigated the dynamics of a single trapped
ion under some certain parity conditions. Our study has shown that
the ZB in the trapped ion is relevant not only to the interference
between different energy components, but also to parity. To
understand the physics related to the realistic Dirac particle, we
have discussed the correspondence of the parity between a
relativistic electron and a trapped ion. Experimental feasibility of
observing our results by a trapped ion has been justified with
currently available techniques. We argue that our study is not only
helpful to explore the ZB effect itself, but also useful to further
understand the quantum characteristic of the ultracold trapped ion.

The work is funded by National Natural Science Foundation of China
under Grants No. 10974225 and No. 11004226, and by Chinese Academy
of Sciences.


\begin{thebibliography}{99}
\bibitem{sch} E. Sch{\"o}dinger, Sitzungsber. Preuss. Akad. Wiss. Phys.
Math. KL \textbf{24}, 418 (1930).

\bibitem{p0} A. O. Barut and A. J. Bracken, Phys. Rev.
D \textbf{23}, 2454 (1981); A.~O.~Barut and W. Tacker, Phys.~Rev.~D. \textbf{31} 1386 (1985).

\bibitem{p6} D.~Hestenes, Found.~Phys. \textbf{20}, 1213 (1990).

\bibitem{p1} V.~A.~Bordovitsyn and I.~M.~Ternov, Russ.~Phys.~J. \textbf{24},
2 (1991).

\bibitem{p4} B.~Thaller, \textit{arXiv}:\textit{quant-ph}/0409079v1 (2004).

\bibitem{p2} P.~Krekora, Q.~Su and R.~Grobe, Phys.~Rev.~Lett. \textbf{93},
043004 (2004).

\bibitem{p5} G.~Sparling Seminaires \& Congres \textbf{4}, 2000 (2007).

\bibitem{p3} Z.~Y.~Wang, and X.~C.~Dong, Phys.~Rev.~A \textbf{77}, 045402
(2008).

\bibitem{ques} K. Huang, Am. J. Phys. \textbf{20}, 479 (1952); N.
Hamdan, A. Altorra and H. A. Salman, Proc. Pakistan Acad. Sci.
\textbf{44}, 263 (2007).

\bibitem{a} L.~Lamata, J.~Leon, T.~Sch\"{a}tz, and E.~Solano,
Phys.~Rev.~Lett. \textbf{98}, 253005 (2007).

\bibitem{s1} W.~Zawadzki, Phys.~Rev.~B \textbf{72}, 085217 (2005).

\bibitem{s2} S. Q. Shen, Phys.~Rev.~Lett. \textbf{95}, 187203 (2005).

\bibitem{s3} J.~Schliemann, D.~Loss, and R.~M.~Westervelt, Phys.~Rev.~Lett.
\textbf{94}, 206801 (2005).
\bibitem{s5} J.~Cserti and G.~D\'{a}vid, Phys. Rev. B \textbf{74}, 172305
(2006).
\bibitem{s4} E.~Bernardes, J.~Schliemann, M.~Lee, J.~C.~Egues, and D.~Loss,
Phys.~Rev.~Lett. \textbf{99}, 076603 (2007).

\bibitem{c} R.~Gerritsma, G.~Kirchmair, F.~Z\"{a}hringer, E.~Solano,
R.~Blatt, and C.~F.~Roos, Nature (Landon) \textbf{463}, 68 (2010).

\bibitem{s7} X. Zhang, Phys. Rev. Lett. \textbf{100}, 113903 (2008).

\bibitem{s8} J. Y. Vaishnav and C. W. Clark, Phys. Rev. Lett. \textbf{100},
153002 (2008).


\bibitem{s9} G. D\'{a}vid and J. Cserti, Phys. Rev. B \textbf{81}, 121417(R)
(2010).

\bibitem{2} http://en.wikipedia.org/wiki/Hermite\_polynomials

\bibitem{explain} For $p=0$ it is easy to prove $\left|\psi_{E_{+}}(p=0)
\right\rangle$ is an even state and $\left|\psi_{E_{-}}(p=0)\right\rangle$
is an odd state, consistent to Eqs. (\ref{eq:+o})-(\ref{eq:-e}).

\bibitem{l} C.~Foudas, \textit{Advanced Particle Physics}, Imperial College.
Lecture 7, (2007).

\bibitem{TLiu} T.~Liu, K.~L.~Wang and M.~Feng, Europhys. Lett. \textbf{86}, 54003 (2009).

\bibitem{cooling} C. Monroe, D. M. Meekhof, B. E. King, S. R. Jefferts, W.
M. Itano, D. J. Wineland, and P. Gould, Phys. Rev. Lett. \textbf{75}, 4011
(1995).

\bibitem{fock} D. M. Meekhof, C. Monroe, B. E. King, W. M. Itano, and D. J.
Wineland, Phys. Rev. Lett. \textbf{76}, 1796 (1996).

\bibitem{cz} J. I. Cirac and P. Zoller, Phys. Rev. Lett. \textbf{74}, 4091
(1995).
\end{thebibliography}
\end{document}